\documentclass[12pt,twoside]{iopart}

\usepackage[bottom]{footmisc} 
\DefineFNsymbols{rjnumber}{{*}{^1}{^2}{^3}{^4}{^5}{^6}{^7}{^8}{^9}
{^{10}}{^{11}}{^{12}}{^{13}}{^{14}}{^{15}}{^{16}}{^{17}}{^{18}}{^{19}}
{^{20}}{^{21}}{^{22}}{^{23}}{^{24}}{^{25}}{^{26}}{^{27}}{^{28}}{^{29}}
{^{30}}{^{31}}{^{32}}{^{33}}{^{34}}{^{35}}{^{36}}{^{37}}{^{38}}{^{39}}{^{40}}} 
\setfnsymbol{rjnumber} 
\usepackage{psfrag}
\usepackage{epsfig}

\newcommand{\eq}[1]{(\ref{#1})}
\newcommand{\fat}[1]{\mbox{\boldmath$#1$}}
\newcommand{\fig}[1]{figure \ref{#1}}
\newcommand{\script}[1]{{\textrm{\scriptsize #1}}}
\newcommand{\su}[1]{_\script{#1}}
\newcommand{\suj}[1]{_{\textrm{\scriptsize #1}}}
\newcommand{\nop}{\hspace{-0.5mm}}

\begin{document}
\title{A covariant formalism of spin precession with respect to a
reference congruence}
\author{Rickard Jonsson\\[2mm]
{\small \it Department of Theoretical Physics,}\\
{\small\it Chalmers University of Technology, 41296 G\"oteborg,
Sweden}
\\[2mm]
\small{\rm E-mail: rico@fy.chalmers.se}
\\[2mm]
\small{\rm
Submitted 2004-12-10, Published 2005-12-08\\
Journal Reference: Class. Quantum Grav. {\bf 23} 37
}
}

\begin{abstract}
We derive an effectively three-dimensional relativistic spin
precession formalism. The formalism is applicable to any spacetime
where an arbitrary timelike reference congruence of worldlines is specified. 
We employ what we call a {\it stopped} spin vector which is the
spin vector that we would get if we momentarily make a pure boost of
the spin vector to stop it relative to the congruence. 
Starting from the Fermi transport equation for the standard
spin vector we derive a corresponding transport equation for the stopped
spin vector. Employing a spacetime transport equation for a vector
along a worldline, corresponding to
spatial parallel transport with respect to the congruence, 
we can write down a precession formula for a
gyroscope relative to the local spatial geometry defined by the congruence. 
This general approach has already been pursued by Jantzen et. al. 
(see e.g. Jantzen, Carini and Bini 1997 {\it Ann. Phys.} {\bf 215} 1), but the algebraic form of our respective expressions differ.
We are also applying the formalism to a novel type of spatial parallel
transport introduced in Jonsson (2006 {\it Class. Quantum Grav.} {\bf 23} 1), as well as verifying the
validity of the intuitive approach of a forthcoming paper 
(Jonsson 2007 {\it Am. Journ. Phys.} {\bf 75} 463)
 where gyroscope precession is explained entirely as a double Thomas
type of effect. We also present the resulting formalism in explicit
three-dimensional form (using the boldface vector notation), and give
examples of applications.
\end{abstract}

\section{Introduction}
In special and general relativity the spin of a gyroscope is
represented by a four-vector $S^\mu$. Assuming that we move the
gyroscope without applying any torque to it (in a system comoving
with the gyroscope), the spin vector will
obey the Fermi transport equation
\begin{eqnarray}\label{rall}
\frac{D S^\mu}{D\tau}= u^\mu \frac{D u^\alpha}{D\tau} S_\alpha
.
\end{eqnarray}
Here $u^\mu$ is the four-velocity of the gyroscope. For a trajectory in a given spacetime, and a spin vector specified at some point
along this trajectory, we can integrate
\eq{rall} to find the spin at any point along the
trajectory. The Fermi transport equation is however deceivingly simple since
we have not inserted explicitly the affine connection coming from the
covariant differentiation. Also, even when we have a flat spacetime and
inertial coordinates (so that the affine connection vanishes)
the equation is more complex than you might think. 
As an example we consider motion
with fixed speed $v$ along a circle in the $xy$-plane, with an
angular frequency $\omega$. Letting the groscope start at $t=0$ at the positive
$x$-axis, we get a set of coupled differential equations 
\begin{eqnarray}\label{rall3}
\frac{d S^x}{dt}&=&\gamma^2 v^2 \omega \sin(\omega t)
\left(S^x\cos(\omega t)+S^y \sin(\omega t) \right) \\
\frac{d S^y}{dt}&=&-\gamma^2 v^2 \omega \cos(\omega t)
\left(S^x\cos(\omega t)+S^y \sin(\omega t) \right) \\
\frac{d S^z}{dt}&=&0, \quad \quad
S^0={\bf v}\cdot{\bf S}
\end{eqnarray}
Here  ${\bf v}=\frac{d{\bf x}}{dt}$ and ${\bf S}$ is the spatial part
of $S^\mu$. For initial conditions
$(S^x,S^y,S^z,S^0)=(S,0,0,0)$ the solutions (see \cite{gravitation} p. 175-176) can
be written as 
\begin{eqnarray}
S^x &=& S \left(\cos[(\gamma-1)\omega t]
+(\gamma-1)\sin[\omega \gamma t] \sin[\omega t]\right) \quad \quad \label{hm1}\\
S^y&=&S \left(\sin[(1-\gamma)\omega t] -(\gamma-1)\sin[\omega \gamma t]\cos[\omega t]\right)  \label{hm2}\\  
S^z&=&0, \quad \quad
S^0=-S R \omega \gamma \sin[\omega \gamma t] \label{hm3}
\end{eqnarray}
Looking at $S^x$ and $S^y$, we note that (written in the particular
form above) the first terms in respective
expression corresponds to a rotation around the $z$-axis, but then
there is also another superimposed rotation with time dependent
amplitude. To find this solution directly from the coupled
differential equations that are the Fermi equations, seems at least at
first sight quite difficult, even for this very symmetric and simple
scenario.

To get a simpler formalism we may consider, not the spin vector
$S^\mu$ itself,
but the spin vector we {\it would} get if we momentarily would stop the
gyroscope (relative to a certain inertial frame) by a pure boost (i.e. a non-rotating boost). This object we will
call the {\it stopped} spin vector. While being a four-vector it is
effectively a three-dimensional object (having zero time component in
the inertial frame in question)
and we will show that the spatial part of this object undergoes pure
rotation with a constant rate for the example of motion along a circle in special relativity.

Knowing that there is a simple algebraic
relation between the stopped and the standard spin vector, the stopped
spin vector can be used as an intermediate step to easily find the standard
spin vector. There is however also a direct physical meaning to the stopped
spin vector, apart from being the spin vector we would get if we
stopped the gyroscope. The stopped spin vector 
{\it directly} gives the spin as perceived in a comoving system, see
section \ref{tttt} for further discussion on this.

\begin{figure}[ht]
  \begin{center}
	\psfrag{A}{$A$}
	\psfrag{B}{$B$}
      	\epsfig{figure=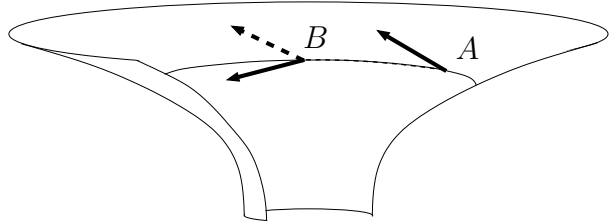,width=8cm,angle=0}
      	\caption{A schematic illustration of how an orbiting gyroscope will
      	precess relative to the spatial geometry of a black
      	hole. The full drawn arrow is the stopped spin vector (stopped
      	with respect to the stationary reference observers) of the gyroscope
      	at two different points along the orbit. The dashed arrow is a vector
      	coinciding with the gyroscope spin vector at $A$ and then
      	parallel transported to $B$ with respect to the spatial geometry. 
	For an intuitive explanation
      	of why the gyroscope precesses relative to the spatial
      	geometry even though there are no torques acting on it, see \cite{intu}.}  
     	\label{fig1}
  \end{center} 
\end{figure}

In this article we will also consider more general reference frames than
inertial ones. For instance we will consider a rotating and accelerating
reference frame. This allows us to apply the formalism, via the equivalence principle,
to describe in a simple three-dimensional manner how a gyroscope
orbiting for instance a rotating black hole will precess relative to
the stationary observers. In \fig{fig1} we illustrate
how a gyroscope spin vector precesses relative to a vector parallel
transported with respect to the spatial geometry.

Given a reference congruence of timelike
worldlines, we first derive a general spacetime
transport equation for the stopped spin vector (stopped relative to the congruence in
question). We then consider a spacetime equation corresponding to {\it spatial} parallel
transport with respect to the spatial geometry defined by the
congruence. For the case of a rigid congruence, we
easily derive such a transport law. Considering a
shearing congruence we use the formalism derived in
\cite{rickinert}.

Having both the transport equation for the stopped spin vector and the
equation for parallel transport, we can put them together and 
thus get an equation for how fast the stopped spin vector 
precesses relative to the local spatial geometry connected to the
reference congruence. As is the case for the inertial congruence, we
will see that the precession corresponds to a simple law of three-rotation.

The general scheme as outlined here has already been pursued by
Jantzen et. al. (see \cite{jantzen}), although the angle of approach and the
algebraic formalisms are different. The explicit use of the
three-dimensional formalism of this paper also appears novel.

This article is complementary to a companion paper \cite{intu}, where
the formalism of relativistic spin precession in three-dimensional
language is derived in a very intuitive manner. This paper verifies,
through a more formal derivation, the result of \cite{intu} for the
particular case of a rigid congruence as assumed in \cite{intu}.

\section{The stopped spin vector}
\begin{figure}[ht]
  \begin{center}
	\psfrag{e}{$\eta^\mu$}
	\psfrag{u}{$u^\mu$}
	\psfrag{S}{$S^\mu$}
	\psfrag{Sb}{$\bar{S}^\mu$}
      	\epsfig{figure=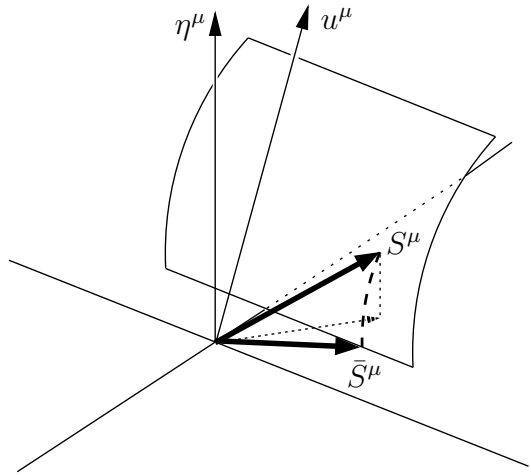,width=7cm,angle=0}
      	\caption{A 2+1 illustration of the relation between the spin
      	vector $S^\mu$ and the stopped spin vector
      	$\bar{S}^\mu$. Through the stopping, the tip of the spin vector
      	can in two dimensions be seen as following the hyperbola 
	connected to the Lorentz transformation down to the local slice. Notice that the 
	stopped spin vector is not in general simply the spatial (projected) part of 
	the standard spin vector (the thin dotted arrow).}  
     	\label{fig2}
  \end{center} 
\end{figure}

Let us denote the local four-velocity of our reference congruence
by $\eta^\mu$. 
We introduce a {\it stopped} spin vector $\bar{S}^\mu$
as the spin vector that we get if we make a pure boost of the spin vector such
that it is at rest with respect to the local congruence line. 
In \fig{fig2} we illustrate in 2+1 dimensions how the two spin vectors
are related to each other.

It follows readily from the Lorentz transformation that we get the
stopped vector by removing the $\eta^\mu$-part of $S^\mu$, and
shortening the part parallel to the spatial direction of motion by a
$\gamma$-factor. Note that the resulting stopped vector is not
in general parallel to the spatial part of $S^\mu$.
Letting $t^\mu$ be a normalized vector orthogonal to $\eta^\mu$
in the direction of motion, we can express the stopped spin vector
as
\begin{eqnarray}\label{stop1}
\bar{S}^\mu&=&\left[{\delta^\mu}_\alpha +\eta^\mu \eta_\alpha
+\left(\frac{1}{\gamma}-1\right)t^\mu t_\alpha \label{s2}\right] S^\alpha
.
\end{eqnarray} 
Here we have adopted the spatial sign convention $(-,+,+,+)$ as we will
throughout the article.
Knowing a little about Thomas precession we may guess that for the
simple case of motion along a circle in an inertial frame as discussed earlier, there is a
simple law of three-dimensional rotation for this object. Indeed in
the following discussion we will show this, and at the same time
consider the effects of rotation coming from having non-inertial
reference frames (connected to $\eta^\mu$).

We also need an explicit expression for the standard spin vector in terms
of the stopped spin-vector $\bar{S}^\mu$. The relationship between the two vectors
follows readily from the Lorentz-transformation: 
\begin{eqnarray}
S^\mu&=&\bar{S}^\alpha {K^\mu}_\alpha \label{s1a} \\
{K^\mu}_\alpha&=&\left[{\delta^\mu}_\alpha + \gamma v \eta^\mu t_\alpha
+(\gamma-1) t^\mu t_\alpha \right]  \label{s1b} 
.
\end{eqnarray}
This we may now insert into the Fermi transport equation to derive an
expression for the {\it stopped} spin vector.

\section{Covariant derivation of the transport equation for the
stopped spin-vector}\label{prima}
In this section we consider gyroscope transport relative to an
arbitrary reference congruence $\eta^\mu$. 
For a spin vector $S^\mu$ transported along a worldline
of four-velocity $u^\mu$,  we have the Fermi transport law
\begin{eqnarray}\label{fermi}
\frac{D S^\mu}{D \tau}=u^\mu S^\rho \frac{D u_\rho}{D \tau}
.
\end{eqnarray}
Using \eq{s1a} in \eq{fermi} readily yields
\begin{eqnarray}
\frac{D\bar{S}^\alpha}{D\tau} {K^\mu}_\alpha=\bar{S}^\alpha \left[
u^\mu {K^\rho}_\alpha \frac{D u_\rho}{D\tau}  -\frac{D{K^\mu}_\alpha}{D\tau}\right]
.
\end{eqnarray}
We need now the inverse of ${K^\mu}_\alpha$ to get an explicit
transport equation for the stopped spin vector. 
Through a general ansatz%
\footnote{We have ${{K^{-1}}^\nu}_\rho
  {K^\rho}_\alpha={\delta^\nu}_\alpha$. The ansatz is of the form
${{K^{-1}}^\nu}_\alpha={\delta^\nu}_\alpha + a t^\mu t_\alpha + b t^\mu
\eta_\alpha + c \eta^\mu t_\alpha +d \eta^\mu \eta_\alpha$.}%
, we find
\begin{eqnarray}\label{kk}
{{K^{-1}}^\nu}_\mu={\delta^\nu}_\mu+\left(\frac{1}{\gamma}
-1\right)t^\nu t_\mu - v \eta^\nu t_\mu
.
\end{eqnarray}
That this is indeed the inverse of ${K^\mu}_\alpha$ is easy to verify%
\footnote{In defining ${K^\mu}_\alpha$ we are free to add terms containing
$\eta_\alpha$, since these anyway die when multiplied by
$\bar{S}^\alpha$. If we instead would have defined 
${K^\mu}_\alpha={\delta^\mu}_\alpha + \frac{1}{\gamma+1} (u^\mu +
\eta^\mu) (u_\alpha - \eta_\alpha)$ 
we would get the inverse 
${{K^{-1}}^\mu}_\alpha={\delta^\mu}_\alpha + \frac{1}{\gamma+1} (u^\mu + \eta^\mu) (\eta_\alpha - u_\alpha)$
. Here the perfect symmetry in $S^\mu,\eta^\mu$ and
 $\bar{S}^\mu,u^\mu$ is transparent. There however does not appear to be
any particular
advantages of this gauge.}.
So we have
\begin{eqnarray}\label{done}
\frac{D\bar{S}^\nu}{D\tau}=\bar{S}^\alpha \left[
u^\mu {K^\rho}_\alpha \frac{D u_\rho}{D\tau}
-\frac{D{K^\mu}_\alpha}{D\tau}\right]{{K^{-1}}^\nu}_\mu
.
\end{eqnarray}
Here we have the desired expression. In \ref{simp} we expand and
simplify this to find
\begin{eqnarray}\label{kott1}
\frac{D \bar{S}^\mu}{D\tau}=&&\frac{\gamma v
}{\gamma+1} \bar{S}^\alpha \left( t^\mu
\left[\frac{D}{D\tau}\left(u_\alpha + \eta_\alpha \right)\right]_\perp 
-t_\alpha \left[ \frac{D}{D\tau}\left(u^\mu + \eta^\mu
\right)\right]_\perp  \right) \nonumber\\&&
+\eta^\mu \bar{S}^\alpha \frac{D\eta_\alpha}{D\tau}
.
\end{eqnarray}
By the perpendicular sign $\perp$ we here mean that we should select
only the part orthogonal to both $t^\mu$ and $\eta^\mu$.
Note that $\frac{D}{D \tau}$ means covariant differentiation along the
gyroscope worldline. Equation \eq{kott1} then tells us how the stopped
spin vector deviates from a parallel transported vector relative to a
freely falling system. 
In fact we notice from the antisymmetric form of
\eq{kott1} that (excepting the $\eta^\mu$ term) it corresponds to a
spatial rotation (see section \ref{gotland} for a more detailed
argument). That seems very reasonable since it insures that the
norm of the stopped spin vector will be constant (consider the
rotation with respect to a freely falling system locally comoving with the
congruence). 
We also see that only if $u^\mu+\eta^\mu$ changes along the gyroscope worldline, with respect to
a freely falling system, do we get a net rotation
relative to this freely falling system. 

Introducing the wedge product defined by $a^\alpha \wedge b^\beta
\equiv a^\alpha  b^\beta - b^\alpha  a^\beta$ and the projection
operator ${P^\mu}_\alpha = {\delta^\mu}_\alpha + \eta^\mu
\eta_\alpha$, we can put \eq{kott1}
in a more compact form
\begin{eqnarray}\label{kottfinal}
{P^\mu}_\alpha\frac{D \bar{S}^\alpha}{D\tau}=\frac{\gamma v
}{\gamma+1} \bar{S}^\alpha \left( t^\mu \wedge
\left[\frac{D}{D\tau}\left(u_\alpha + \eta_\alpha \right) \right]_\perp \right)
.
\end{eqnarray}
Incidentally we may note that, as regards $t^\mu$-components within the bracketed
expression, we do not need the $\perp$ sign. Any $t^\mu$ components within the bracketed
expressions will cancel due to the anti-symmetrization as is easy to
see. We however keep the $\perp$ sign to indicate orthogonality to $\eta^\mu$.
The simple form of \eq{kottfinal} appears to be novel.

\section{Application to flat spacetime, and inertial congruences}
While we have yet to put the formalism in its final form, some
applications and discussion may be useful already at this
point for the simple case of an
inertial reference congruence in special relativity.

\subsection{Employing the spatial curvature of the gyroscope trajectory}
As a particular example, consider a flat spacetime with an inertial
congruence. For this case it is not hard to show, see e.g
\cite{rickinert}, that the spatial curvature of a trajectory depends on the
four-acceleration as
\begin{eqnarray}
\left[\frac{Du_\alpha}{D\tau}\right]_\perp=\gamma^2 v^2\frac{n_\alpha}{R}
.
\end{eqnarray}
Here $R$ is the spatial curvature%
\footnote{As is illustrated in \cite{rickinert} there are plenty of
ways to define spatial curvature measures in general, but for an inertial congruence
most of these coincide with the standard projected curvature that we
here assume.}
and $n^\mu$ is a normalized
four-vector, orthogonal to the inertial congruence $\eta^\mu$,
pointing in the direction of spatial curvature.
Using this in \eq{kottfinal} we get
\begin{eqnarray}\label{kottfinal2}
{P^\mu}_\alpha\frac{D \bar{S}^\alpha}{D\tau}=\gamma v
(\gamma-1) \bar{S}^\alpha \left( t^\mu \wedge \frac{n_\alpha}{R} \right)
.
\end{eqnarray}
As we will see in the following section, this differential equation
corresponds to a three-dimensional rotation.

\subsection{Three-dimensional formalism, for flat spacetime and an inertial congruence}\label{gotland}
Choosing inertial coordinates adapted to the inertial congruence in question so that
$\bar{S}^\mu=(0,{\bf \bar{S}})$, $t^\mu=(0,{\bf \hat{t}})$ and
$n^\mu=(0,{\bf \hat{n}})$ we get from \eq{kottfinal2}
\begin{eqnarray}\label{e1}
\frac{d{\bf \bar{S}}}{d\tau}=\gamma v (\gamma-1)\left[{\bf \hat{t}}({\bf
\bar{S}} \cdot \frac{{\bf \hat{n}}}{R} ) -\frac{{\bf \hat{n}}}{R} ({\bf \bar{S}}
\cdot {\bf \hat{t}} )  \right]
.
\end{eqnarray}
The expression within the brackets is a vector
triple product and we may write it as a double cross
product. Letting ${\bf v}=v {\bf \hat{t}}$ we get
\begin{eqnarray}\label{e2}
\frac{d{\bf \bar{S}}}{d\tau}=\gamma 
(\gamma-1) \left(\frac{{\bf \hat{n}}}{R}\times {\bf v}\right) \times {\bf \bar{S}} 
.
\end{eqnarray}
Rather than using $\tau$ we could use local time $\tau_0$ (the time as
experienced by observers at rest relative to the inertial congruence
in question) in which
case we get a gamma factor less on the right hand side.
\begin{eqnarray}\label{aye}
\frac{d{\bf \bar{S}}}{d\tau_0}=(\gamma-1) \left(\frac{{\bf \hat{n}}}{R}\times {\bf v}\right) \times {\bf \bar{S}} 
.
\end{eqnarray}
This is the famous Thomas precession, in stopped spin vector
three-formalism. Introducing $\fat{\Omega}$ as the precession
vector, around which the stopped spin vector rotates, we can alternatively write \eq{aye} as
\begin{eqnarray}
\frac{d{\bf \bar{S}}}{d\tau_0}&=&\fat{\Omega} \times {\bf \bar{S}}   \label{ayea} \\
\fat{\Omega}&=&(\gamma-1) \left(\frac{{\bf \hat{n}}}{R}\times {\bf v}\right).\label{ayeb}
\end{eqnarray}
Looking at \eq{ayea} component-wise, it is a set of coupled
differential equations, just like the standard Fermi equations.
Unlike the Fermi-equations however, the new equations
correspond to a simple law of rotation (precession).

\subsection{The circular motion revisited}
As a specific example we may consider, as in the introduction, the
precession of a gyroscope transported at constant
speed $v$ around a circle of radius $R$ in the $z=0$ plane. 
Assuming a motion with a clockwise angular velocity $\omega=v/R$, the
counterclockwise angular velocity $\Omega$ for the precession of the stopped spin vector
is then according to \eq{ayeb} given by
\begin{eqnarray}\label{how}
\Omega=(\gamma-1)\omega.
\end{eqnarray}
Consider then for instance the net precession after one lap.
The local time per lap is simply $2 \pi/\omega$ and hence the net precession
angle (in radians) around the plane normal is given by $2\pi
(\gamma-1)$. If the circular motion is counter-clockwise, the
precession is clockwise and vise versa. 

\subsection{Re-deriving the solution for the standard spin vector}\label{revhep}
We can also trivially find the solution for the standard (projected)
spin vector for the case of circular motion with constant speed
with initial conditions as listed in the example in the introduction. 
We know that the standard (projected) spin vector
is related to the stopped spin vector through a lengthening of the stopped spin vector in the forward
direction of motion ${\bf \hat{t}}$  by a $\gamma$-factor. We have then
\begin{eqnarray}\label{hpp}
{\bf S}={\bf \bar{S}}+(\gamma-1)({\bf \bar{S}}\cdot{\bf \hat{t}}){\bf \hat{t}}.
\end{eqnarray}
Using the notation of the previous subsection we have then trivially for the case at hand
\begin{eqnarray}
{\bf \bar{S}}&=&S \cos(\Omega t ) {\bf \hat{x}} -S \sin(\Omega t) {\bf \hat{y}}\\
{\bf \hat{t}}&=&-\sin(\omega t) {\bf \hat{x}} + \cos(\omega t) {\bf \hat{y}}.
\end{eqnarray}
Using these expressions in \eq{hpp} we immediately get the desired
solution. Using elementary rules for manipulating
the trigonometric functions we can write it in the
form of \eq{hm1}-\eq{hm3}. If we
are interested also in $S^0$, it is given by the orthogonality of the
standard spin vector and the four-velocity as $S^0={\bf S}\cdot{\bf
  v}$. Note that by use of the stopped spin vector formalism there is
effectively no differential equation solving involved for this simple case.

\subsection{A special relativistic theorem of spin precession 
for planar constant velocity motion}
For motion in a circle with constant velocity, the Fermi equation can
be solved without 
use of the stopped spin vector formalism%
, although the solution is a bit complicated. What about if we
consider motion with constant velocity along some other curve, say a
part of a parabola or some more irregular curve? 
Then the Fermi equation would likely appear to be very complicated to
solve analytically in the general case. Using the method with the
stopped spin vector the solution can however trivially be found for arbitrary
curves. First let us state a small theorem that we will then easily prove.

\begin{quote}
{\it The stopped spin vector of a gyroscope transported with constant 
speed $v$ along a smooth curve in a spatial plane 
in a flat spacetime will rotate a net clockwise angle 
around the normal of the plane given by
$\Delta \alpha\su{precess}=(\gamma-1) \Delta \alpha\su{curve}$ where
$\Delta \alpha\su{curve}$ is the net counterclockwise turning angle of
the tangent direction of the curve.}
\end{quote}
Note that the parameter $\Delta \alpha\su{curve}$ may be larger than
$2\pi$. For a simple closed curve (one that is not crossing itself),
assuming the gyroscope to be transported once around the curve, we
have $\Delta \alpha\su{curve}=2\pi$.

This theorem is easily proven by dividing an arbitrary smooth curve into
infinitesimal segments 
within which we can consider
the local curvature radius to be constant.
Letting $\omega$ denote the counter-clockwise angular velocity of the
forward direction of motion ${\bf \hat{t}}$ (so $\omega=d\alpha\su{curve}/dt$),
we have according to \eq{how} the clockwise angular velocity as
$\Omega=(\gamma-1)\omega$. Thus the net angles
of the gyroscope precession and the turning of the forward direction
of the curve, along the segment in question, are related through 
$d\alpha\su{precess}=(\gamma-1)d\alpha\su{curve}$.
Adding up the precession contributions from all the segments of
the curve we get
\begin{eqnarray}\label{hh}
\Delta \alpha\su{precess}=(\gamma-1)\Delta \alpha\su{curve}.
\end{eqnarray}
Thus the theorem is proven. Note that while the motion is assumed to 
be in a plane, the spin vector may point off the plane.

\subsection{Some consequences of the theorem}\label{concon}
\begin{figure}[ht]
  \begin{center}
	\psfrag{a}{Initial projected spin vector}
	\psfrag{aa}{Final projected spin vector}
      	\epsfig{figure=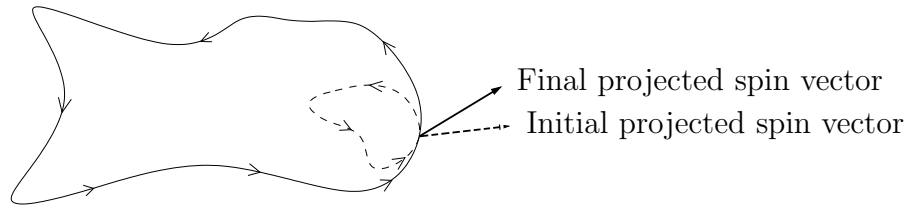,width=7cm,angle=0}
      	\caption{Illustrating that for a fixed initial direction of motion, 
	fixed initial 
	spin vector, and fixed constant speed $v$
	-- the final spin vector after one lap around 
	any simple smooth closed curve is independent of the shape of the curve.}
     	\label{xf1}
  \end{center} 
\end{figure}

We can draw a conclusion from the above proven theorem (also knowing that 
there is a simple algebraic relation between the stopped and the 
standard spin-vector) that can be expressed in terms of the 
standard spin vector, without reference to the stopped spin vector.
Consider then a smooth simple closed curve and let a certain point along 
this curve be the initial position for the gyroscope. For given
initial spin vector, initial direction of motion%
\footnote{One cannot in general keep the standard (unlike the stopped) 
spin vector fixed while altering the initial direction of motion of 
the gyroscope since 
the standard spin vector must be orthogonal to the gyroscope
four-velocity.}
and constant speed $v$, the final spin vector (after one lap around the loop) 
is independent of the shape of the loop as illustrated in \fig{xf1}.

But of course, the theorem is stronger than this. Given an arbitrary, 
not necessarily closed but smooth curve along which we transport the 
gyroscope with constant velocity, we can trivially find the standard spin 
vector at any point along the curve. We take the spatial part of the 
initial spin vector and shorten the part parallel to the 
direction of motion by a $\gamma$-factor to form the initial stopped 
spin vector. For any given curve ${\bf x}(\lambda)$ we then calculate 
the initial direction of the curve together with the direction of the 
curve at the point in question. Then, modulo a winding number times $2 
\pi$%
\footnote{The only non-trivial part of calculating the turning angle 
lies in finding out the number 
of turns taken by the curve since for a curve ${\bf x}(\lambda)$ we only get 
the turning angle $\Delta \alpha\su{curve}$ up to a term $2\pi n$, 
where $n$ is an integer, from the local quantity $\frac{d{\bf 
x}}{d\lambda}$},
we can trivially find the corresponding $\Delta \alpha\su{curve}$ 
and thus through \eq{hh} the corresponding stopped spin vector at the point in 
question. Lengthening the parallel part of the stopped spin vector by 
a factor $\gamma$, we get the spatial part of the standard spin vector 
at the point in question. If we are interested in the zeroth component of the standard spin 
vector it is given by $S^0={\bf S} \cdot {\bf v}$. 
Thus solving a possibly very complicated differential equation 
is reduced to performing a few algebraic steps%
\footnote{Again modulo the winding number mentioned earlier. For many 
cases, like for instance for a parabola, this however presents no problem at all.}. 

\subsection{More complicated motion}
For motion in a plane where the velocity is not constant, the procedure is analogous to that 
described in section \ref{concon} except that we need to integrate (a single 
integral which may or may not be complicated to solve analytically) to find $d\alpha\su{precess}$.
For the most general motion, not necessarily confined to a 
plane and with a speed that may vary, it is however not just a matter of
ordinary integration%
\footnote{One could for instance represent a finite precession (rotation)
  by a vector whose direction determines the axis of rotation and
  whose norm determines the angle (in radians) of the precession. It is
  however easy to realize that for a a finite such rotation (like the
  net rotation after some finite stretch along a trajectory) followed
  by an infinitesimal rotation around some other axis -- one cannot in
  general simply add the two corresponding rotation vectors 
  (to first order) to form a new rotation vector. 
  Of course there are examples of
  non-planar motion, like motion along a helix for instance,  where the
  precession vector remains in the same direction for which case it is
  a simple matter of integration after all to find the net rotation of
  the stopped spin vector.}.
Given an arbitrary motion ${\bf x}(\tau_0)$ along a smooth curve we
  can however solve a differential equation, given by \eq{ayea} and \eq{ayeb},
for ${\bf \bar{S}}$.
Likely this differential equation will be simpler to solve than the Fermi equation.

\subsection{A comment on the relation between the intrinsic angular momentum,
  the projected spin vector, the gyroscope axis and the 
stopped spin vector}\label{spax}
To gain further intuition on the meaning of the stopped spin vector it
may be useful to explore how it is related to other vectors of physical
interest connected to the gyroscope spin. In particular we may consider
the gyroscope intrinsic angular momentum, and the momentary 
direction of the gyroscope axis as perceived in the reference 
system in question (where the observers are integral curves of 
$\eta^\mu$). 

\begin{figure}[ht]
  \begin{center}
	\psfrag{a}{\small The gyroscope axis direction}
	\psfrag{aa}{\small The stopped spin vector}
	\psfrag{aaa}{\small The projected spin vector}
	\psfrag{aaaa}{\small The standard spin vector}
	\psfrag{aaaaa}{\small The worldsheet of the gyroscope axis}
      	\epsfig{figure=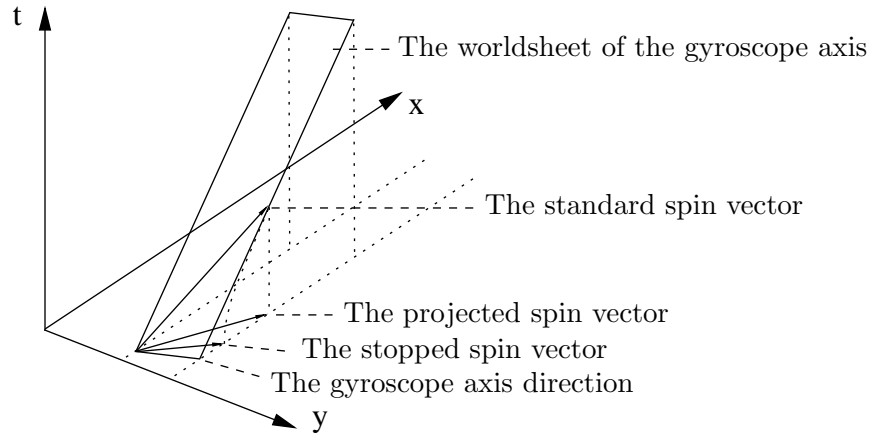,width=7cm,angle=0}
      	\caption{A sketch in 2+1 dimensions of vectors related to
      	a spinning gyroscope.}
     	\label{xf2}
  \end{center} 
\end{figure}

Consider then a gyroscope moving along a straight line in the
$xy$-plane in special relativity (using inertial coordinates) 
with constant speed. The  
gyroscope axis is assumed to lie in the plane of motion and to be tilted somewhere between the forward and the sideways 
direction. In 2+1 dimensions we can easily visualize the worldsheet 
of the gyroscope central axis
as well as various vectors of interest, see \fig{xf2}.

\begin{figure}[ht]
  \begin{center}
	\psfrag{a}{$\frac{1}{\gamma}$}
	\psfrag{aa}{$1$}
	\psfrag{aaa}{$\gamma$}
	\psfrag{b}{${\bf \hat{t}}$}
	\psfrag{x}{${\bf X}$}
	\psfrag{sb}{${\bf \bar{S}}$}
	\psfrag{s}{${\bf S}$}
	\psfrag{sl}{${\bf S}\su{L}$}
      	\epsfig{figure=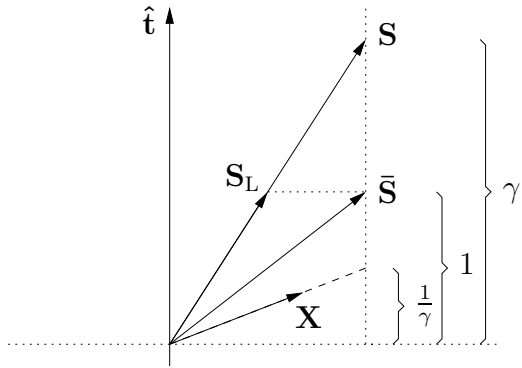,width=7cm,angle=0}
	\caption{The three different directions in question are simply related 
      	through a stretching by a gamma factor in the direction of 
      	motion. In this illustration a gamma factor of 2 was assumed,
      	with motion in the upwards direction (${\bf \hat{t}}$).
	Note that the depicted norm of the gyroscope axis vector ${\bf X}$ is arbitrary.}
     	\label{xf3}
  \end{center} 
\end{figure}

We note that there are (at least) three different spatial directions
of relevance for the gyroscope. It is 
easy to realize (length contraction) that the direction of the gyroscope axis is simply related to the 
direction of the stopped spin vector through a gamma factor. Given any of these directions the other two can thus easily be found.
Furthermore one can show, at least for an idealized scenario as considered
in \ref{spinapp}, that the the intrinsic angular momentum, that
we will denote ${\bf S}\su{L}$, is in fact
given by ${\bf S}/\gamma$. The various vectors involved are illustrated in \fig{xf3}.

\subsection{Four vectors, four differential equations}
Consider a spatial vector ${\bf X}$ that connects the base of
the gyroscope to the tip of the gyroscope, as perceived in the
reference system connected to $\eta^\mu$. We understand that this vector
evolves according to a simple rule of rotation given by \eq{aye}
modulated by a contraction by a factor of $\gamma$ in the direction of motion.
It is a short exercise to show that this means that ${\bf X}$ in fact
obeys a rather compact differential equation
\begin{eqnarray}\label{ldiff}
\frac{d{\bf X}}{d\tau_0}=-\gamma^2 \frac{d {\bf v}}{d\tau_0} \left[
  {\bf X}\cdot {\bf v}\right].
\end{eqnarray}
We can perform a corresponding analysis for the projected spin vector
to find%
\footnote{This also follows readily from the standard Fermi equations
  for the case of inertial coordinates in special relativity.} 
\begin{eqnarray}\label{rall5}
\frac{d{\bf S}}{d\tau_0}=\gamma^2 {\bf v} \left[ {\bf S} \cdot \frac{d {\bf
  v}}{d\tau_0}\right].
\end{eqnarray}
The equations for the stopped spin vector can be written in the form 
\begin{eqnarray}\label{rall6}
\frac{d{\bf \bar{S}}}{d\tau_0}=\frac{\gamma-1}{v^2}\left(\frac{d {\bf
  v}}{d\tau_0} \times {\bf v} \right)\times {\bf \bar{S}}.
\end{eqnarray}
From \eq{rall5}, letting ${\bf S}=\gamma {\bf S}\su{L}$, 
we readily find 
\begin{eqnarray}\label{rall7}
\frac{d{\bf S}\su{L}}{d\tau_0}=\gamma^2 {\bf v} \left[ {\bf S}\su{L} \cdot \frac{d {\bf
  v}}{d\tau_0}\right] - \gamma^2 v \frac{dv}{dt} {\bf S}\su{L}.
\end{eqnarray}
Comparing the four differential equations we see that they are all
quite compact, although the equation for the stopped spin vector,
corresponding to a pure rotation, is more likely to be simple to solve
(as we have seen for the example of motion on a circle).

\subsection{A comment on the meaning and purpose of the stopped spin vector}\label{tttt}
One might argue that the object of physical interest is the intrinsic (spin) 
angular momentum of the gyroscope which is given by ${\bf
  S}/\gamma$, or perhaps the observed direction of the gyroscope central axis.
From this point of view the stopped spin vector is in a sense a means to an
end. By using the stopped spin vector as an intermediate step we can find the solutions to
otherwise quite complicated differential equations for the objects of
physical interest.
From a mathematical point of view this is certainly sufficient
to motivate the use of the stopped spin vector. 
There is however more to the stopped spin vector than this.
In particular we note that the stopped spin vector {\it directly} gives us the spin as perceived in a 
comoving system. For instance, if the stopped spin
vector is at a $45^\circ$ angle with respect to the forward direction -- 
so it will be with respect to a system comoving with the
gyroscope%
\footnote{If the stopped spin vector has certain components with
  respect to a set of base vectors adapted to the reference
  congruence in question,  then those components precisely corresponds
  to the components of the standard spin vector with respect to a
  boosted version (a pure boost to comove with the gyroscope) of the
  base vectors just mentioned. This viewpoint is mentioned in
  \cite{gravitation} p. 1117, although they do not consider general
  spacetimes and velocities.}.
This is contrary to the standard spin vector which only
gives the spin direction with respect to the comoving system  
after a Lorentz transformation.
Consider the following example. A
gyroscope is suspended inside a
satellite such that no torque is exerted on the gyroscope as seen from
the satellite. The satellite is assumed to be orbiting along some predetermined
smooth simple closed curve, on a plane in special relativity\footnote{The
  general argument works  also for gyroscopes orbiting the earth in
a general relativistic treatment. More on this in section \ref{axi}.}, 
using it's jet engines to stay on the path.
Suppose then that we wish to measure, from the
satellite, the precession angle of the
gyroscope (as predicted by relativity) after a full orbit (or maybe several full orbits). 
We note that the direction of the gyroscope relative to the satellite itself is
not a good measure%
\footnote{The satellite may have had an initial rotation
from the start or the jet-engines may give it one. Also, even if it
would have zero proper rotation then the gyroscope would keep its
direction relative to the satellite and thus would not turn at all
relative to the satellite.}. 
Assuming that we have a couple of fixed stars, we can however use the
direction of these stars (as
perceived from the satellite) as guidelines to set up a reference
system within the satellite%
\footnote{We also assume that the satellite has some way
of knowing when it is at its initial position (so it knows when to
calibrate its coordinates with respect to the stars).}
For this scenario the stopped spin vector is exactly the physical
object that we are interested in.
It exactly represents the gyroscope
direction relative to the star-calibrated reference system of the
satellite. Thus if the stopped spin vector turns a certain angle, that
is precisely the turning angle of the gyroscope relative to the star-calibrated
reference system of the satellite. 

While we are here focusing on spinning gyroscopes, it should also be noted
 that the formalism of the stopped spin vector is immediately
 applicable  to describe the resulting rotation of {\it any} object
 which has zero {\it proper} (comoving) rotation.

In conclusion, the stopped spin vector may be used as an intermediate
step to simplify the calculation of the evolution of the intrinsic angular
momentum (spin) of a gyroscope, or the perceived direction of the
gyroscope axis. The stopped spin vector is however 
also of direct physical
importance since it gives us the spin as perceived in a comoving
system.

So far we have only given examples that apply to flat spacetime, 
and inertial reference frames. As we will see in the following sections the 
stopped spin vector can be just as useful also for curved spacetimes
and non-inertial reference frames.

\section{Spatial parallel transport} \label{spat}
The transport equation \eq{kottfinal} tells us how the stopped spin vector
deviates from a vector that is parallel transported with respect to
the spacetime geometry. This by itself is however not really what we are
after if the reference congruence is non-inertial. To get a
truly three-dimensional formalism, we in stead want an expression telling us how
fast the stopped spin vector deviates (rotates) from a vector that is parallel
transported with respect to the {\it spatial} geometry determined by
the congruence. As is demonstrated in \cite{jantzen} and in \cite{rickinert}, it is possible derive a {\it spacetime}
transport law corresponding to a {\it spatial} parallel transport.
For the simple, and perhaps most useful, case of a rigid congruence%
\footnote{The congruence may rotate and accelerate but it may not shear or expand.}
the issue is sufficiently simple that we will briefly review it in the coming subsection.

\subsection{Rigid congruence}
Suppose then that we have a rigid congruence with nonzero acceleration
$a^\mu$, nonzero rotation tensor  ${\omega^\mu}_\nu$ but with
vanishing expansion-shear tensor ${\theta^\mu}_\nu$%
\footnote{The kinematical invariants of the congruence are defined as
  (see \cite{gravitation} p. 566): 
The expansion scalar $\theta=\nabla_\alpha \eta^\alpha$, the
  acceleration vector
$a^\mu=\eta^\alpha \nabla_\alpha \eta^\mu$,
the shear tensor
$\sigma_{\mu \nu}=\frac{1}{2} \left(\nabla_\rho \eta_\mu {P^\rho}_\nu +
\nabla_\rho \eta_\nu {P^\rho}_\mu \right) -\frac{1}{3} \theta P_{\mu
  \nu}$ and the rotation tensor
$\omega_{\mu \nu}=\frac{1}{2} \left({P^\rho}_\nu \nabla_\rho
\eta_\mu - {P^\rho}_\mu \nabla_\rho \eta_\nu   \right)$. Furthermore
  we employ what we denote the expansion-shear tensor 
$\theta_{\mu \nu}=\frac{1}{2} \left({P^\rho}_\nu \nabla_\rho
\eta_\mu + {P^\rho}_\mu \nabla_\rho \eta_\nu   \right)$.}.

\begin{figure}[ht]
  \begin{center}
      	\epsfig{figure=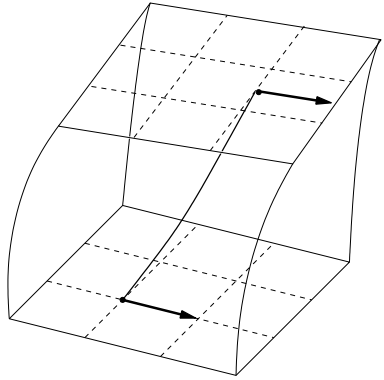,width=5cm,angle=0}
      	\caption{A 2+1 illustration of transporting a spatial vector
      	along a worldline, seen from freely falling coordinates
      	locally comoving with the congruence. As the reference coordinates
      	rotate due to ${\omega^\mu}_\alpha$, so
      	should the vector in order for it to be proper spatially
      	transported.}  
     	\label{fig3}
  \end{center} 
\end{figure}

In \fig{fig3} we show an illustration of the spacetime transport of a
vector orthogonal to the congruence.

It is easy to show that in the coordinates of a freely falling system
$(t,x^k)$, locally comoving with the congruence, the
velocity of the congruence points (assuming vanishing
${\theta^\mu}_\nu$) is to first order given by
\begin{eqnarray}
v^k={\omega^k}_j x^j + a^k t
.
\end{eqnarray}
Knowing that the velocity of the congruence is zero to lowest order,
relative to the inertial system in question,
we need not worry about length contraction and such. It is then easy
to realize that the proper spacetime transport law of a spatial vector $k^\mu$
corresponding to standard spatial parallel transport is
\begin{eqnarray}\label{init}
\frac{D k^\mu}{D\tau}= \gamma {\omega^\mu}_\alpha k^\alpha + b \eta^\mu
.
\end{eqnarray}
Here $b$ can easily be determined from the orthogonality of $k^\mu$
and $\eta^\mu$\footnote{From the orthogonality $k^\mu \eta_\mu=0$
follows (differentiate $\frac{D}{D\tau}$ along the gyroscope
worldline) that $\frac{D k^\mu}{D \tau}\eta_\mu =-k^\mu
\frac{D\eta_\mu}{D\tau}$. Contracting both sides of \eq{init} by
$\eta_\mu$ gives $b=k^\mu \frac{D \eta_\mu}{D \tau}$.}. Here we have then a spacetime transport equation
corresponding to spatial parallel transport, for the case of a
non-shearing (non-expanding) congruence.

\subsection{Including shear and expansion}
For a more complicated congruence that is shearing and expanding, it
is not quite so obvious how to define the spatial parallel
transport. Indeed as discussed in e.g \cite{jantzen} and
\cite{rickinert}, there are several ways of doing this. 
We will here
follow the approach of \cite{rickinert}, and consider two different such
parallel transports. These transports are connected to two
different ways of defining a spatial curvature for a test particle
worldline, with respect to the congruence
\begin{eqnarray}
\textrm{Projected:}\quad&&
\frac{1}{\gamma^2} \left[\frac{Du^\mu}{D\tau}\right]_\perp
= [a^\mu]_\perp+2v ({\omega^\mu}_\alpha t^\alpha +[{\theta^\mu}_\alpha t^\alpha]_\perp)   + v^2
\frac{n\su{ps}^\mu}{R\su{ps}} \\[3mm]
\textrm{New:}&&
\frac{1}{\gamma^2} \left[\frac{Du^\mu}{D\tau}\right]_\perp
= [a^\mu]_\perp+2v {\omega^\mu}_\alpha t^\alpha + v^2
\frac{n\su{ns}^\mu}{R\su{ns}}  
.
\end{eqnarray}
Here $R\su{ps}$ and $n^\mu\su{ps}$ are
the curvature and the curvature direction that we get if we project the the spacetime trajectory
down along the congruence onto a local timeslice (orthogonal to the
congruence at the point in question). The suffix 'ps' stands for 'Projected Straight'. 
The curvature $R\su{ns}$ and the curvature direction $n^\mu\su{ns}$ are
defined with respect to deviations from a certain (new) notion of a
spatially straight line. 
The latter is defined as a line that with
respect to variations in the projected curvature, leaves the
integrated spatial distance (as defined by the congruence) unaltered (to first order in the
variation). As it turns out, a straight line with respect to
this definition, has in general a non-zero projected curvature when
the congruence is shearing.  The suffix 'ns' stands for 'New-Straight'. This particular curvature 
is connected to Fermat's principle, and optical
geometry \cite{rickinert,genopt}.

For brevity we let the suffix 's' denote either 'ps', or 'ns'. 
Introducing $C\su{ps}=1$, $C\su{ns}=0$ we can then
express both curvatures jointly as
\begin{eqnarray}\label{jcurve}
\frac{1}{\gamma^2} \left[\frac{Du^\mu}{D\tau}\right]_\perp
&=& [a^\mu]_\perp+2v ({\omega^\mu}_\alpha t^\alpha +C\su{s}[{\theta^\mu}_\alpha t^\alpha]_\perp)   + v^2
\frac{n\su{s}^\mu}{R\su{s}}.
\end{eqnarray}
From these two curvature measures one can introduce corresponding
equations for spatial parallel transports \cite{rickinert}. A joint
expression for the parallel transport of a vector $k^\mu$ is given by
\begin{eqnarray}\label{ps}
\displaystyle \frac{Dk^\mu}{D\tau} = \gamma k^\alpha {\omega^\mu}_\alpha
 + \gamma (2 C\su{s}-1 ) k^\alpha ({\theta^\mu}_\beta t^\beta \wedge
 t_\alpha) + \eta^\mu k^\alpha \frac{D\eta_\alpha}{D\tau} 
.
\end{eqnarray}
Here $\frac{D\eta_\alpha}{D\tau}$ is the covariant derivative along
the (gyroscope) worldline in question.
Notice that for vanishing shear expansion tensor, the two transports
both correspond to \eq{init}.

Having defined two types of parallel transport according to \eq{ps},
we can define corresponding 
covariant differentiations  along a curve as
\begin{eqnarray}
\frac{D\su{s} k^\mu}{D\su{s} \tau} =
 \frac{D k^\mu}{D\tau} - \gamma
 k^\alpha \left( {\omega^\mu}_\alpha 
 + (2 C\su{s} -1)({\theta^\mu}_\beta t^\beta \wedge t_\alpha) \right)
 - \eta^\mu k^\alpha \frac{D\eta_\alpha}{D\tau}  \label{transproj} 
.
\end{eqnarray}
These derivatives then tells us how fast a vector deviates from a
corresponding parallel transported vector (momentarily parallel to the
vector in question). Substituting $k^\mu \rightarrow \bar{S}^\mu$ and
using \eq{kottfinal} we get the equations for how fast the stopped spin
vector precesses relative to a spatially parallel transported vector
(of the two types). First we however rewrite \eq{kottfinal}.

\section{Rewriting the stopped spin vector transport equation}
We saw in the preceding section how the kinematical invariants of the
congruence entered naturally in the definition of spatial parallel
transport. 
We can also expand  $(\frac{D\eta_\alpha}{D\tau} +
\frac{D u_\alpha}{D\tau})$, in the transport equation \eq{kottfinal}
for the stopped spin vector, in terms of the kinematical invariants of
the congruence.
First of all we have
\begin{eqnarray}
\frac{D\eta_\alpha}{D\tau}&=&u^\rho \nabla_\rho \eta_\alpha=\gamma ( \eta^\rho +  v t^\rho)  \nabla_\rho \label{katta}
\eta_\alpha
.
\end{eqnarray}
Also we know that (see e.g \cite{gravitation} p. 566)
\begin{eqnarray}
\nabla_\rho \eta_\alpha=\omega_{\alpha \rho}+\theta_{\alpha \rho}-a_\alpha \eta_\rho
.
\end{eqnarray}
Using \eq{katta}, we
have then
\begin{eqnarray}\label{deta}
\frac{D\eta_\alpha}{D\tau}=\gamma v \left(\omega_{\alpha \rho}t^\rho 
+\theta_{\alpha \rho} t^\rho    \right) + \gamma a_\alpha
.
\end{eqnarray}
Using this together with \eq{jcurve} in \eq{kottfinal}, also adding
the proper $\eta^\mu$-term enabling the removal of the projection
operator in \eq{kottfinal}, we readily find
\begin{eqnarray}\label{delettps}
\frac{D \bar{S}^\mu}{D\tau}=\frac{\gamma v
}{\gamma+1} \bar{S}^\alpha t^\mu \wedge&&\hspace{-0mm} \Bigg[
\gamma(\gamma+1)a_\alpha +
\gamma v (2\gamma+1)\omega_{\alpha \rho} t^\rho \nonumber\\&&
+\gamma v (2\gamma C\su{s}+1) \theta_{\alpha \rho} t^\rho 
+\gamma^2 v^2 \frac{n_{\script{s}\alpha}}{R\su{s}} 
\Bigg]
+\eta^\mu \bar{S}^\alpha \frac{D\eta_\alpha}{D\tau}
.
\end{eqnarray}
Notice that we have omitted the perpendicular signs ($\perp$) on
$\theta_{\alpha \rho}t^\rho$ and $a_\alpha$ since these objects are already
orthogonal to $\eta^\mu$ and any $t^\mu$ components die due to the anti-symmetrization.

\section{The rotation of the stopped spin vector relative to a parallel
transported vector}
Now it is time to put together the results of the preceding two
sections. What we want is the net rotation of the stopped spin vector
relative to a spatially parallel transported vector. 
Using  \eq{delettps} and \eq{transproj} (setting
$k^\alpha=\bar{S}^\alpha$), 
we then readily find
\begin{eqnarray}\label{finp}
\frac{D\su{s} \bar{S}^\mu}{D\su{s} \tau}=\bar{S}^\alpha&&
\Bigg[ \gamma^2 v (t^\mu \wedge a_{\alpha})  +
(\gamma-1)(2\gamma+1) (t^\mu \wedge \omega_{\alpha \rho} t^\rho)-\gamma
{\omega^\mu}_\alpha \nonumber\\&&\hspace{-0mm}+(2\gamma^2
C\su{s}-1) (t^\mu \wedge \theta_{\alpha \rho} t^\rho) +\gamma v
(\gamma -1) \left(t^\mu \wedge \frac{n_{\script{s} \alpha}}{R\su{s}} \right)
\Bigg]
.
\end{eqnarray}
Here $C\su{ps}=1$ and $C\su{ns}=0$. So this gives
us how fast a gyroscope stopped spin vector deviates from a corresponding
(spatially) parallel transported vector. In particular considering the expression
in a freely falling system locally comoving with the congruence, we
understand that the expression within the brackets on the right hand
side is simply the effective rotation tensor relative to the spatial
geometry.

It could be practical with an expression corresponding to \eq{finp}
but where the proper four-acceleration is explicit. 
Using \eq{kottfinal}, \eq{transproj} and  \eq{deta} we readily find 
\begin{eqnarray}\label{fin2p}
\frac{D\su{s}
\bar{S}^\mu}{D\su{s}\tau} \nop=\nop
&&
\frac{\gamma
v}{\gamma \nop +\nop 1} \bar{S}^\alpha
t^\mu \wedge \left[ \left[\frac{D u_\alpha}{D\tau}\right]_\perp \nop \nop \nop \nop
  + \nop \gamma
a_\alpha \nop + \nop \gamma v \omega_{\alpha \rho}t^\rho \nop + \nop
(2\gamma C\su{s} \nop-\nop 1)\frac{\gamma \nop + \nop 1}{\gamma
v}\theta_{\alpha \rho}t^\rho\right]
\nonumber\\ &&\hspace{-0mm}
-\gamma {\omega^\mu}_\alpha \bar{S}^\alpha
.
\end{eqnarray}
Notice that the expression for the four-acceleration here (naturally)
is independent of what curvature measure that we use. Still \eq{fin2p} 
depends on what curvature measure we are using (manifesting itself in
the occurrence of $C\su{s}$) assuming non-zero $\theta_{\alpha \rho}t^\rho$, since the transport
laws for the two types of spatial parallel transport differs.

\section{Three-dimensional formalism, assuming rigid congruence}
We can rewrite \eq{finp} and \eq{fin2p} as purely three-dimensional equations. For any specific global labeling of the congruence lines (i.e. any specific
set of spatial coordinates adapted to the congruence) we can locally
choose a time slice orthogonal to the congruence so that $\bar{S}^\mu=(0,{\bf
\bar{S}})$. This then uniquely defines the three-vector ${\bf \bar{S}}$ at any
point along the gyroscope trajectory. Analogous to what we did in
going from \eq{kottfinal2} to \eq{e2}, for a set of vectors $\bar{S}^\mu$,
$t^\mu$ and $k^\mu$ orthogonal to the congruence, we let
$\bar{S}^\alpha t^\mu\wedge k_\alpha \rightarrow \left({\bf k} \times
{\bf \hat{t}} \right) \times {\bf \bar{S}}$.%
\footnote{Strictly speaking, what we mean by the cross product  ${\bf a}
\times {\bf b}$ of two three-vectors ${\bf a}$ and ${\bf b}$ is
$[\textrm{Det}(g_{ij})]^{-\frac{1}{2}} \epsilon^{ijk}a_j b_k$ where the indices have been lowered with the
local three-metric (again assuming local coordinates orthogonal to the
congruence). Notice that in general (for congruences with rotation)
there are no global time-slices that are orthogonal to the
congruence. The local three-metric corresponding to local orthogonal
coordinates is however well defined everywhere anyway. For a shearing
(expanding) congruence it will however be time dependent (whatever
global time slices we choose).}
Also we let ${\omega^\mu}_\alpha t^\alpha   \rightarrow \fat{\omega} \times
\fat{\hat{t}}$%
\footnote{Letting $\omega^\mu=(0,\fat{\omega})$ in coordinates locally
  comoving with the congruence, we have $\omega^\mu=\frac{1}{2}
  \frac{1}{\sqrt{g}} \eta_\sigma \epsilon^{\sigma \mu \gamma \rho}
  \omega_{\gamma \rho}$, where $g=-\textrm{Det}[g_{\alpha
  \beta}]$ and $\epsilon^{\sigma \mu \gamma \rho}$ is $+1$, $-1$ or
  $0$ for $\sigma \mu \gamma
  \rho$  being an even, odd or no permutation of $0,1,2,3$ respectively.}. 
For simplicity, let us assume that the congruence has
vanishing shear and expansion.%
\footnote{This incidentally implies that the 'orthogonal' three-metric
mentioned in a previous footnote is time independent.}
 For this case the two different
approaches to spatial curvature radius coincide and we will drop any
instances of subscripts 'ns' or 'ps'. Introduce then 
${\bf a}\suj{gyro}=\frac{d^2 {\bf x}}{d\tau_0^2}$, where ${\bf x}$ and $\tau_0$
are the inertial coordinates of a system locally comoving with the
congruence%
\footnote{Working in another set of spatial coordinates ${\bf
a}\suj{gyro}$ naturally transforms as a three-vector.}%
.  Also denoting the acceleration of the reference congruence relative to an
inertial system locally comoving with the reference congruence by ${\bf
a}\suj{ref}$, we get from \eq{fin2p}\footnote{
Notice that $D/D\tau$ corresponds to covariant differentiation
with respect to the three-metric.}
\begin{eqnarray}\label{fin3}
\frac{D \bar{\fat{S}}}{D\tau}
=\frac{\gamma^3
}{\gamma+1} \left( \left[{\bf a}\suj{gyro}+\frac{1}{\gamma} ({\bf
a}\suj{ref}+\fat{\omega}\times{\bf v})\right]\times {\bf v} \right) \times {\bf
\bar{S}}-\gamma \fat{\omega} \times {\bf \bar{S}}
.
\end{eqnarray}
This is a perfect match with the result of the intuitive derivation
performed in \cite{intu}.

Analogously we may study \eq{finp} for the particular case of vanishing shear, thus
considering a rigid congruence. The three-dimensional version of
this equation then becomes
\begin{eqnarray}\label{fin4}
\frac{D\bar{\bf S}}{D\tau}= 
\Bigg[&&\hspace{-0mm} \gamma^2 v ({\bf a}\suj{ref} \times {\bf \hat{t}}) - \gamma \fat{\omega} +
(\gamma-1)(2\gamma+1)(\fat{\omega} \times {\bf \hat{t}})\times{\bf
\hat{t}} 
\nonumber\\&&\hspace{-0mm}
+ \gamma v(\gamma -1) \left(\frac{\bf \hat{n}}{R}\times {\bf \hat{t}} \right)\Bigg] \times \bar{\bf S} 
.
\end{eqnarray}
We may simplify this expression a bit by introducing
$\fat{\omega}=\fat{\omega}_\parallel+\fat{\omega}_\perp$, where
$\parallel$ and $\perp$ means parallel respectively perpendicular to
${\bf \hat{t}}$. Also using ${\bf v}=v {\bf \hat{t}}$ we readily find
\begin{eqnarray}\label{fin5}
\frac{D \bar{\bf S}}{D \tau}= 
\Bigg[&&\hspace{-0mm}\gamma^2 ({\bf a}\suj{ref} \times {\bf v}) - \gamma \left( \fat{\omega}_\parallel +
\left(2\gamma-\frac{1}{\gamma}\right)\fat{\omega}_\perp   \right) 
\nonumber\\&&\hspace{-0mm}
+ \gamma  (\gamma -1) \left(\frac{\bf \hat{n}}{R}\times {\bf v} \right)\Bigg] \times \bar{\bf S} 
.
\end{eqnarray}
Again this is a perfect match with the intuitive formalism of \cite{intu}. 

\subsection{The rotation vector relative to the reference observers.}
Letting $\tau_0=\gamma d\tau$ denote local time for an observer
comoving with the congruence we can write \eq{fin3} and \eq{fin5} respectively as
\begin{eqnarray}\label{rjx1}
\frac{D \bar{\bf S}}{D \tau_0}= {\bf \Omega} \times \bar{\bf S}.
\end{eqnarray}
Here ${\bf \Omega}$ is given by \eq{fin3}
\begin{eqnarray}\label{rjx3}
\fat{\Omega}=\frac{\gamma^2}{\gamma+1} ({\bf a}\su{gyro} \times {\bf
  v}) +\frac{\gamma}{\gamma+1}({\bf a}\suj{ref} \times {\bf v})
  -\fat{\omega}_\parallel -\left(2-\frac{1}{\gamma} \right) \fat{\omega}_\perp  
.
\end{eqnarray}
This form is practical if the gyroscope is freely falling, in
which case ${\bf a}\suj{gyro}=0$. Alternatively we can get ${\bf \Omega}$ from \eq{fin5}
\begin{eqnarray}\label{rjx2}
{\bf \Omega}=\gamma ({\bf a}\suj{ref} \times {\bf v}) - \fat{\omega}_\parallel -
\left(2\gamma-\frac{1}{\gamma}\right)\fat{\omega}_\perp    
+ (\gamma -1) \left(\frac{\bf \hat{n}}{R}\times {\bf v} \right) 
.
\end{eqnarray}
This form is practical if the gyroscope follows some
predetermined path while being acted on by forces.

\subsection{A note on the gyroscope axis and the projected spin vector}\label{addiff}
From the simple relation (see section \ref{spax}) between the stopped
spin vector and the projected spin vector and the gyroscope axis
respectively, we can use the law of rotation for the stopped spin
vector to derive corresponding differential equations for ${\bf S}$
and {\bf X} 
\begin{eqnarray}\label{finalcombo}
\frac{d {\bf S}}{d\tau_0}&=&\gamma^2 {\bf v} \left[{\bf S} \cdot \frac{d {\bf
      v}}{d\tau_0} \right]+
\fat{\Omega}_{\script{e} \parallel} \times [{\bf S}]_\perp +
\fat{\Omega}_{\script{e} \perp} \times \left(
\frac{1}{\gamma}  [{\bf S}]_\parallel + \gamma [{\bf S}]_\perp
      \right) \label{combo1}\\
\frac{d {\bf X}}{d\tau_0}\nop\nop\nop\nop \nop &=&\nop\nop\nop\nop -\gamma^2 \frac{d {\bf v}}{d\tau_0}
      \left[{\bf X} \cdot {\bf v} \right]+
\fat{\Omega}_{\script{e} \parallel} \times [{\bf X}]_\perp +
\fat{\Omega}_{\script{e} \perp} \times \left(
 \gamma [{\bf X}]_\parallel + \frac{1}{\gamma} [{\bf X}]_\perp
      \right). \label{combo2}
\end{eqnarray}
Here we have for brevity introduced 
\begin{eqnarray}
{\bf \Omega}\su{e}=\gamma ({\bf a}\suj{ref} \times {\bf v}) - \fat{\omega}_\parallel -
\left(2\gamma-\frac{1}{\gamma}\right)\fat{\omega}_\perp.
\end{eqnarray}
Note that the $\frac{d {\bf v}}{d\tau_0}$ entering \eq{combo1} and \eq{combo2} is the velocity derivative relative to the
reference frame (not relative to a freely falling frame). 
We note that these differential equations are more complicated than
the ones for the stopped spin vector. We conclude that if we are interested in ${\bf S}$ or ${\bf X}$, it is
likely wise to first solve the equation for the stopped spin vector
and then (as in section \ref{revhep}) use the result to find ${\bf S}$ or ${\bf X}$.

\section{Motion along a straight line in static geometry}
\begin{figure}[ht]
  \begin{center}
    \psfrag{t=0}{$t=0$}
    \psfrag{t=dt}{$t=dt$}
    \psfrag{g}{${\bf g}$}
    \psfrag{v}{$v$}
    \epsfig{figure=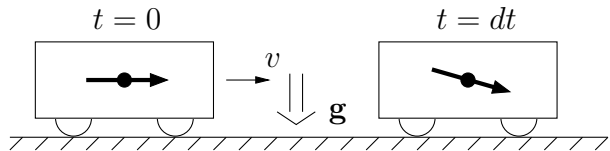,width=8cm,angle=0}
    \caption{A train moving relative to a straight platform with
      proper upward acceleration. A gyroscope with a torque free
      suspension will precess clockwise (for positive $v$).} 
    \label{fig12}
  \end{center} 
\end{figure}

As a first example of how one may use the derived formalism, consider a train moving along a straight
spatial line in some static geometry. In fact, to be specific, we
may consider the train to be moving relative to an upwards
accelerating platform in special relativity. 
On the train we have
suspended a gyroscope so that there are no torques acting on it in the
comoving system. We thus consider gyroscope motion along a straight
line, with respect to a non-rotating but accelerating reference frame. Letting
${\bf g}=-{\bf a}_{\script{ref}}$ and $\tau_0=\gamma \tau$, \eq{fin5} is immediately reduced to
\begin{eqnarray}\label{perfa}
\frac{d{\bf \bar{S}}}{d\tau_0}=-\gamma 
\left( {\bf g}\times {\bf v} \right) \times {\bf \bar{S}}.
\end{eqnarray}
We understand that a gyroscope initially pointing in the forward
direction will tip forward as depicted in \fig{fig12}.

Note that
the stopped spin vector with respect to the
platform corresponds precisely to the spin vector as perceived
relative to the train. For example, if the stopped spin vector points $45^\circ$
  down from the horizontal direction, the gyroscope as seen from the
  train points $45^\circ$ down from the horizontal direction.
To express the gyroscope precession with respect
to coordinates comoving with the train we therefore just let 
$\tau_0 \rightarrow \gamma \tau$ in \eq{perfa} and we have the
precession explicitly in terms of the time $\tau$ on the train.
Relative to the train, the gyrocope thus precesses at a steady rate
given by $\Omega\su{relative train}=\gamma^2 v g$.
This means that the train in fact has a proper rotation, but more on
this is given in \cite{intu}.

We can parameterize the gyroscope trajectory by
the distance $s$ along the platform rather than the time $\tau_0$. Then \eq{perfa} can
be expressed as
\begin{eqnarray}
\frac{d{\bf \bar{S}}}{ds}= -\gamma \left( {\bf g}\times {\bf \hat{t}} \right) \times {\bf \bar{S}}   
.
\end{eqnarray}
Assuming the train velocity to be low, the tipping angle per distance traveled
is thus independent of the velocity. We have simply
$d\alpha/ds \simeq g$. Thus on a stretch of length $\delta s$ we get a net
rotation $\delta \alpha$
\begin{eqnarray}
\delta \alpha \simeq g {} \delta s
.
\end{eqnarray}
If we want to express $\delta s$ and $g$ and in SI units we must divide
the right hand side by $c^2$ (expressed in SI units). 
Setting $\delta s=10^3 \textrm{ m}$ and $g=9.81 \textrm{ m/s}^2$ we get simply
\begin{eqnarray}
\delta \alpha=\frac{9.81 \cdot 10^3}{(3 \cdot 10^8)^2} \approx 1\cdot
10^{-13} {} ({\textrm{rad}})
.
\end{eqnarray}
This is quite a small angle, and we understand that the relativistic
effects of gyroscope precession for most cases here at Earth are
small. Notice how simple this calculation was in the
stopped spin vector three-formalism. 

\section{Axisymmetric spatial geometries, and effective rotation
vectors}\label{axi}
The equations \eq{fin3} and \eq{fin5} both describe how the
gyroscope rotates with respect to a coordinate frame that is parallel
transported with respect to the spatial geometry.
Suppose then that we consider
motion in the equatorial plane of some axisymmetric geometry. As a specific example we may want
to know the net rotation of the gyroscope relative to its initial
configuration after a full orbit (not necessarily a circular orbit). 
We must then take into consideration
that a parallel transported frame in general will be rotated relative
to its initial configuration after a complete orbit. 

We can however introduce a new reference frame, that rotates relative to local
coordinates spanned by ${\bf \hat{r}}$ and $\fat{\hat{\varphi}}$, in the
same manner as a parallel transported reference frame does on a plane.
In other words, if we
for instance  consider a counterclockwise displacement ($\delta\varphi$,$\delta r$), then relative to the local
vectors ${\bf \hat{r}}$ and $\delta \fat{\hat{\varphi}}$, the new reference
frame should turn precisely $\delta \varphi$ clockwise. Such a
reference frame would always return to its initial configuration after
a full orbit.

To find the rotation of the new reference frame with respect to a
parallel transported frame, we first investigate how a vector that is
parallel transported with respect to the curved axisymmetric geometry
rotates relative to the local coordinates spanned by ${\bf \hat{r}}$ and $\fat{\hat{\varphi}}$.

\subsection{The rotation of a parallel transported vector relative
  to ${\bf \hat{r}}$ and $\fat{\hat{\varphi}}$} 
\begin{figure}[ht]
  \begin{center}
    \psfrag{rdrp}{$(r+\delta r)\delta \varphi$}
    \psfrag{rp}{$r\delta \varphi$}
    \psfrag{b}{$\alpha+\delta\alpha$}
    \psfrag{a}{$\alpha$}
    \psfrag{R}{$R_0$}
    \psfrag{ds}{$\delta s$}
    \psfrag{da}{$\delta \alpha$}
    \epsfig{figure=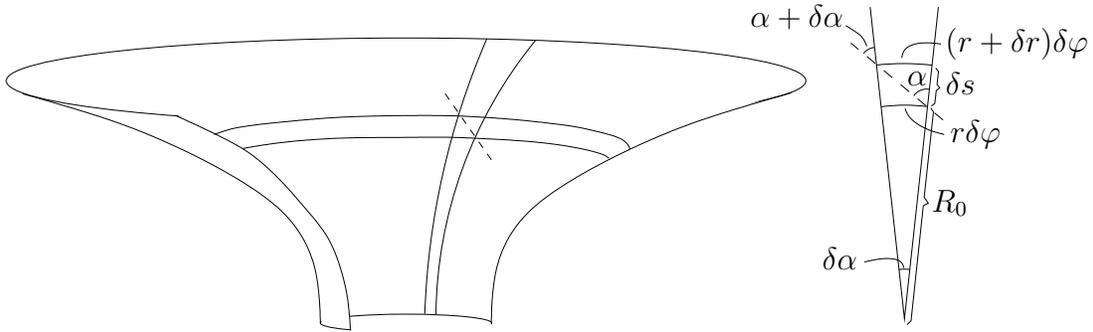,width=13cm,angle=0}
	\caption{Cutting out a section of a certain $d\varphi$ and $dr$ of the embedded geometry (to the
      	left) and putting it on a flat plane (to the right). Note that
      	$r$ is the circumferential radius, and $R_0$ is the radius of
      	curvature for a circle at the $r$ in question (not to be
      	confused with the $R$ of the trajectory along which we are
      	parallel transporting the vector)}  
     	\label{fig14}
  \end{center} 
\end{figure}

The line element for a two-dimensional axisymmetric spatial geometry can be written in the form%
\footnote{Note that if we consider for instance a Kerr black hole,
where we (in standard representation) have $d\varphi dt$-terms, we cannot
simply select the spatial terms (without $dt$) to get the spatial line
element. There is however an effective spatial geometry also for this case. We may derive
the form of this geometry by for instance sending photons back and
forth between nearby spatial points and checking the proper time that passes.}
\begin{eqnarray}
ds^2=g_{rr} dr^2 + r^2 d\varphi^2
.
\end{eqnarray}
As depicted in \fig{fig14} we can imagine an embedding of the
geometry, where we cut out a small section and put it on a flat
plane. What we want is an expression for how much a vector that is parallel
transported, for example along the depicted straight dashed line,  rotates relative to the local
coordinates ${\bf \hat{r}}$ and $\fat{\hat{\varphi}}$. We understand
that the rotation angle corresponds to the angle $\delta \alpha$ as depicted.

\noindent
Using the notations introduced in \fig{fig14} we have simply
\begin{eqnarray}
\hspace{1.25cm} R_0 \delta \alpha &&= r \delta \varphi \label{pr1}\\
(R_0+ds)\delta \alpha &&= (r+dr) \delta \varphi  \label{pr2}
.
\end{eqnarray}
Eliminating $R_0$ and using $ds=\sqrt{g_{rr}} dr$ it follows
readily that
\begin{eqnarray}\label{rdone}
\delta\alpha=\frac{\delta\varphi}{\sqrt{g_{rr}}}.
\end{eqnarray}
So this tells us how a parallel transported vector turns relative to the local ${\bf \hat{r}}$ and
$\fat{\hat{\varphi}}$, for a small displacement in $\varphi$ and $r$.

\subsection{The new reference frame, and the effective rotation vector}
On a flat plane, the corresponding expression to \eq{rdone} is of course simply
\begin{eqnarray}\label{rdonex}
\delta\alpha=\delta \varphi.
\end{eqnarray}
A reference frame that with respect to ${\bf \hat{r}}$ and
$\fat{\hat{\varphi}}$ rotates as if we had a flat geometry would then
according to \eq{rdone} and \eq{rdonex} rotate relative to
a parallel transported reference frame with an
angular frequency (never mind the sign for the moment) 
\begin{eqnarray}
\omega\su{space}&=&\frac{d\varphi}{d\tau_0}
\left(\frac{1}{\sqrt{g_{rr}}} - 1\right).
\end{eqnarray}
Note also that we have
\begin{eqnarray}
\left|\frac{d\varphi}{d\tau_0}\right|=\left|\frac{r d\varphi}{d\tau_0}\right| \frac{1}{r}
=\frac{1}{r}|{\bf v} \cdot \fat{\hat{\varphi}}|
=\frac{1}{r}|{\bf v} \times {\bf \hat{r}}|.
\end{eqnarray}
Thinking about the sign for a second, we realize that with respect to
the 'would-be-flat' reference frame, a parallel transported reference
frame will have a rotation given by
\begin{eqnarray}\label{katt}
\fat{\omega}\su{space}=\frac{1}{r}\left(\frac{1}{\sqrt{g_{rr}}}-1\right) {\bf
  v} \times {\bf \hat{r}}.
\end{eqnarray}
Knowing that infinitesimal
rotations can simply be added (to lowest order), using \eq{katt}
together with \eq{rjx2}
 and letting ${\bf g}=-{\bf a}\suj{ref}$, we get the gyroscope rotation relative to the
'would-be-flat'-grid as%
\footnote{If the geometry in question contains regions where
  the circumferential radius has a minimum (in 2 dimensions one
  may call these regions {\it necks} from the appearance of an
  embedding of such regions), one can modify \eq{katt2} a
  little by introducing a $\pm$ sign in the $\frac{1}{\sqrt{g_{rr}}}$-term. If ${\bf \hat{r}}$, which by
  definition is taken to point away from the center of symmetry,
  points in the direction of increasing $r$, we choose the positive
  sign, otherwise the negative sign should be chosen. Note that
  $\frac{1}{\sqrt{g_{rr}}}$ is zero for radii where the sign changes, so there is no discontinuity in $\fat{\Omega}_{\script{effective}}$.}
\begin{eqnarray}\label{katt2}
\fat{\Omega}_{\script{effective}}=&&
(\gamma-1)\left(\frac{{\bf \hat{n}}}{R} \times {\bf v} \right) 
-\gamma ({\bf g} \times {\bf v})
\nonumber 
- \fat{\omega}_\parallel-\fat{\omega}_\perp
\left(2\gamma-\frac{1}{\gamma} \right)  \\
&&
+\frac{1}{r} \left(\frac{1}{\sqrt{g_{rr}}}-1
\right) {\bf v} \times {\bf \hat{r}} 
.
\end{eqnarray} 
We can integrate this equation to find the net precession
of a gyroscope transported along any path in the axisymmetric geometry.

\subsection{Comments on the integrability}
As a particular application of \eq{katt2}, we can consider the
net precession of a gyroscope transported along some closed orbit.
Since the 'would-be-flat' reference frame returns to its original
configuration after a full turn, we just integrate the effects of the infinitesimal
rotations following from \eq{katt2} to calculate the net turn. 
Notice however that to do this straightforwardly, we need
$\fat{\Omega}_{\script{effective}}$ in the coordinate base of the
reference frame (i.e. the would-be-flat frame). In general we however only have
$\fat{\Omega}_{\script{effective}}$ in the coordinates adapted to the
stationary observers. For most cases where we would be interested in
motion in an axisymmetric geometry, like motion in the
equatorial plane of a Kerr black hole, this however presents no
problem. Then all rotations are in the plane of motion and the
rotation vector $\fat{\Omega}_{\script{effective}}$ is constant (in
the $\hat{z}$-direction) in the coordinate basis of the reference frame.
Notice that the $\tau_0$ implicitly entering these equations in the $\fat{\Omega}_{\script{effective}}$  is the
proper time for a stationary observer. If we instead want to express the
precession in global (Schwarzschild) time, we just multiply by the time
dilation factor.

Even assuming all rotations to be in the plane of motion, we must
still in general integrate to get the net precession of the gyroscope%
\footnote{Parameterizing the trajectory by some parameter $\lambda$, we
  understand that time dilation, $R$, $\fat{\omega}$, $v$ and ${\bf
  g}$ all depends on $\lambda$. Assuming all rotations to be in the
  plane of motion it is effectively a single (scalar)
  integral (of the net rotation angle around the $z$-axis).}.
For the particular
  case of circular motion with constant speed, assuming the time
  dilation (i.e. the lapse),
  $\fat{\omega}$ and ${\bf g} \cdot {\bf \hat{r}}$ to be independent of
  $\varphi$ (as is the case for the equatorial plane of a Kerr black hole), there is however no need to
  integrate at all since all the terms of \eq{katt2} are constant. The result
  follows immediately, assuming that we know
  $\fat{\omega}$, $\fat{g}$ and $g_{rr}$. Incidentally it
  follows from \eq{pr1} and \eq{pr2} that $R=r \sqrt{g_{rr}}$ for
  circular motion.

\subsection{Comment regarding $\fat{g}$, $\fat{\omega}$ and $g_{rr}$}
The reference background (fixed to the stars) around a spinning
planet, like the Earth, is both accelerating and curved. Also there is frame dragging due
to the planet rotation, giving a non-zero rotation of the stationary reference observers. If we have the spacetime
metric, we can easily find $\fat{\omega}$, $\fat{g}$ and $g_{rr}$.
If we do not have an exact spacetime metric however, as is the case
for a spinning planet, we need some approximate method (like the
Post-Newtonian approximation) to estimate $\fat{\omega}$, $\fat{g}$
and $g_{rr}$. Once this is done, assuming the approximation to be
valid, \eq{katt2} gives an accurate description of the precession even considering
relativistic speeds.

In the case of a rotating (Kerr) black hole, we do know the
metric, and the precession relative to the stationary observers can
readily be calculated. Notice that within the ergosphere 
, there are no stationary
(timelike) observers. Still we can in principle use
the formalism of this paper also within the ergosphere. To do this we
consider coordinates that rigidly rotate around the black hole
sufficiently fast to be timelike in the region in question. Indeed
for the particular case of circular motion there is a paper
\cite{rindper}, that uses this technique.

\subsection{Free orbit at large radii from a Schwarzschild
black hole}
As a simple example, consider a freely falling gyroscope (${\bf a}_{\script{gyro}}=0$)
orbiting in the equatorial plane of a Schwarzschild black hole. Using the static
observers as our reference congruence, we have $\fat{\omega}=0$. Then it follows from
\eq{rjx3} that we have
\begin{eqnarray}\label{yesrotny}
\fat{\Omega}_{\script{effective}}=\frac{\gamma}{\gamma+1}{\bf a}_{\script{ref}} \times {\bf v} + \fat{\omega}_\script{space}
.
\end{eqnarray}
The Schwarzschild line element in the equatorial plane is given by
\begin{eqnarray}
d\tau^2=\left(1+2\phi\right)dt^2 -\left(1+2\phi\right)^{-1}dr^2 -r^2
d\varphi^2 
.
\end{eqnarray}
Here $\phi=-\frac{M}{r}$. For convenience we now consider large $r$, so that $M/r$ is small. We have then
the acceleration of the freely falling frames $g\simeq\frac{\partial
\phi}{\partial r}$ (counted positive in the inwards direction). It follows readily, using \eq{katt}, that for this case we have
\begin{eqnarray}
\fat{\omega}_{\script{space}}&=& \frac{1}{r} \left(\sqrt{1+2 \phi}-1
\right)  {\bf v} \times {\bf \hat{r}} \\
&\simeq&\frac{\phi}{r} \cdot {\bf v} \times {\bf \hat{r}}\\ 
&\simeq&-g \cdot {\bf  v} \times {\bf \hat{r}}\label{ospace}
.
\end{eqnarray}
For the large $r$ in question the velocities are low and we may set $\gamma \simeq 1$. Using \eq{ospace} together with ${\bf a}_{\textrm {ref}}=-{\bf g}$ and ${\bf g}=-g {\bf \hat{r}}$ in \eq{yesrotny} gives
\begin{eqnarray}
\fat{\Omega}_{\script{effective}}&\simeq&
-\frac{1}{2} ({\bf g} \times {\bf v}) 
- {\bf g} \times {\bf v} \\ \label{h3}
&=&-\frac{3}{2} ({\bf g} \times {\bf v}) 
.
\end{eqnarray}
This result was derived by W. de Sitter in 1916 (although in a quite
different manner than that described here, see \cite{gravitation} p. 1119). 
We may note that one third of the net effect comes from the spatial
geometry. Using a little bit of Newtonian mechanics it is easy
to derive that for a satellite orbiting the Earth at a radius $R
\simeq R_{\script{Earth}}
$,
inserting the proper factor of $c$ to handle SI-units, \eq{h3}
amounts to
\begin{eqnarray}\label{h4}
\Omega_{\script{effective}}=&&
\frac{3}{2c^2} \frac{GM}{R^2} 
\sqrt{\frac{GM}{R} } \simeq 1.3 \cdot 10^{-12} \textrm{rad/s}
\nonumber\\&&\hspace{-5mm}
\simeq 4.0 \cdot 10^{-5}\frac{\textrm{rad}}{\textrm{year}}
\simeq 8.3 \frac{\textrm{arcsec}}{\textrm{year}} 
.
\end{eqnarray}
Knowing that the exterior metric of the Earth is approximately
Schwarzschild, we have then an approximation of the effective rotation vector
for a gyroscope orbiting the Earth. We can refine this
approximation by considering an appropriate non-zero $\fat{\omega}$,
as discussed earlier. Note that, as discussed in section \ref{tttt},
the derived precession is the precession with respect to a
star-calibrated reference system on the satellite.

In \cite{gravitation} p. 1117-1120, a similar explicitly three-dimensional
formalism of spin precession is derived. It is only valid in the
Post-Newtonian regime however. The precession given by \eq{katt2} is
however exact (assuming an ideal gyroscope). For instance, considering the above example of free circular motion in a static geometry, we can easily calculate the exact expressions for $g$ and $v$, and thus express the gyroscope precession rate arbitrarily close to the horizon.

\section{Summary and conclusion}
We have seen how we in a covariant manner can derive an effectively 
three-dimensional spin precession formalism in a general spacetime. In
particular the simple form of \eq{kottfinal} seems novel.

In \cite{gravitation} p. 1117 a similar approach is taken where they consider
only the standard spin vector, but expressed relative to a
{\it boosted} set of base vectors. They however only apply it to the post-Newtonian regime. 

As mentioned earlier, Jantzen et. al. (\cite{jantzen,
  jantzen2,jantzen2app}) have already pursued the same general idea,
  although the specific approach and final form of the results
  differ. In particular they have not employed the explicit 3-dimensional formalism. 

While the general formalism is derived assuming a general congruence, 
it seems to have its greatest use as a simple
three-dimensional formalism assuming a non-shearing congruence. Then
we have a fixed spatial geometry and the spatial parallel transport is
unambiguous. For this particular case, the derived three-dimensional
formalism verifies the result of the intuitive derivation of
\cite{intu}.  We have also given examples of how the three dimensional formalism can be
used to easily find results of physical interest.

\appendix

\section{Simplifying \eq{done}}\label{simp}
In the expansion of the second term within the brackets of \eq{done}
there will according to \eq{kk} be terms of the type
$\frac{Dt^\mu}{D\tau}$. These can be rewritten in terms of
$\frac{Du^\mu}{D\tau}$ and $\frac{D\eta^\mu}{D\tau}$ since we have
\begin{eqnarray}
u^\mu=\gamma(\eta^\mu+v t^\mu) \quad \quad  \rightarrow \quad \quad
t^\mu=\frac{u^\mu}{\gamma v}-\frac{\eta^\mu}{v}
.
\end{eqnarray}
Dealing with $\frac{Du^\mu}{D\tau}$ rather than $\frac{Dt^\mu}{D\tau}$
is convenient since the former readily can be expressed in terms of spatial
curvature and velocity changes relative to the congruence, see
\cite{rickinert}. Also $\frac{Du^\mu}{D\tau}$ has a direct
physical relevance. 
Using the identity $\frac{\gamma-1}{\gamma v}=\frac{\gamma v}{\gamma+1}$, it
is then easy to derive an alternative form of ${K^\mu}_\alpha$
\begin{eqnarray}
{K^\mu}_\alpha={\delta^\mu}_\alpha + \frac{1}{\gamma+1} 
\left(u_\alpha-\gamma \eta_\alpha\right)
\left(\eta^\mu + u^\mu\right)
.
\end{eqnarray}
Using this in the second term within the brackets of \eq{done} we have
\begin{eqnarray}\label{come}
\frac{D{K^\mu}_\alpha}{D\tau}&=&\frac{D}{D\tau}\left[ \frac{1}{\gamma+1} 
\underbrace{\left(u_\alpha-\gamma \eta_\alpha\right)}_{\gamma v t_\alpha}
\underbrace{\left(\eta^\mu + u^\mu\right)}_{(\gamma+1)\eta^\mu+\gamma v
t^\mu} \right]
.
\end{eqnarray}
As we expand this expression there will be terms containing 
$\eta^\mu$, $\eta_\alpha$ and $t^\mu t_\alpha$. These we will
disregard for the following reasons. Terms containing $\eta_\alpha$
will anyway die when multiplied by $\bar{S}^\alpha$ (as they are in \eq{done}). Terms containing
$\eta^\mu$ we will disregard since we for the moment only are interested in
${P^\mu}_\alpha \bar{S}^\alpha=({g^\mu}_\alpha + \eta^\mu \eta_\alpha)
\bar{S}^\alpha$.  When we have a neat expression for this
we can find the $\eta^\mu$-part a posteriori using the orthogonality of
$\bar{S}^\alpha$ and $\eta_\alpha$. We will disregard terms of the
type $t^\mu t_\alpha$ since we know that these must cancel anyway for
$\bar{S}^\alpha$ to stay normalized (as we know it must by
construction of the Fermi transport and the relation to the stopped
spin vector)%
\footnote{From normalization follows that 
$\bar{S}_\alpha \frac{D \bar{S}^\alpha}{D \tau}=0$. For the particular case where
$\bar{S}^\alpha=\bar{S} t^\alpha$ momentarily, it follows that any
net term of the form $a t^\mu t_\alpha$ in the right hand side
of \eq{done} must vanish. Since the
parameter $a$ does not depend on $S^\alpha$ it follows that it must vanish
entirely. 
The point is that the form of \eq{come} is such that, when expanded
it can be written as a sum of tensors of the
type $A^\mu B_\alpha$. Letting the suffix $\perp$ indicate that only
the part orthogonal to both $\eta^\mu$ and $t^\mu$ should be selected,
each such term can
be written in the form $(t^\mu t_\rho A^\rho + [A^\mu]_\perp)(t_\alpha
t^\sigma B_\sigma+ [B_\alpha]_\perp)$. Adding up the resulting terms of the
type $t^\mu t_\alpha$ (including the terms of this type coming from
the first term within the brackets of \eq{done}) into a single term $a  t^\mu t_\alpha$ we
{\it know} that $a$ must be zero.}.
Note however that in principle, we should contract with the inverse
${{K^{-1}}^\nu}_\mu$%
\footnote{
Note from \eq{kk} that the effect of contracting ${{K^{-1}}^\mu}_\alpha$ 
with a contravariant vector is that it shortens the $t^\mu$-component of the 
vector by a $\gamma$-factor, while the rest of the on-slice
(orthogonal to $\eta^\mu$) part 
of thee vector is unaffected.}%
, before disregarding the terms of the described
types (see \eq{done}). The form of the inverse is however such that we can carry out
the effective cancellations prior to applying the inverse%
\footnote{If the inverse had contained for instance terms
of the type $t^\nu \eta_\mu$ -- we could not cancel $\eta^\mu$
terms directly within the brackets of \eq{done}. That the inverse is not containing any such terms is a benefit
of the particular gauge choice in choosing ${K^\nu}_\mu$ -- where we
had a freedom to include any terms containing $\eta_\mu$.}.
We then readily find
\begin{eqnarray}\label{part1}
\frac{D{K^\mu}_\alpha}{D\tau}&\stackrel{\textrm{eff}}{=}&
\frac{\gamma v}{\gamma +1}
\left( 
\left[\frac{D u_\alpha}{D\tau} - 
\gamma \frac{D \eta_\alpha}{D\tau}\right]_\perp t^\mu+
t_\alpha
\left[\frac{D u^\mu}{D\tau} + \frac{D \eta^\mu}{D\tau} \right]_\perp
\right)
.
\end{eqnarray}
By the perpendicular sign $\perp$ we here mean that we should select
only the part orthogonal to both $t^\mu$ and $\eta^\mu$. By
$\stackrel{\textrm{eff}}{=}$ we indicate that the equality holds
excepting terms of the type $\eta^\mu$, $\eta_\alpha$ and $t^\mu t_\alpha$.
In an analogous manner we readily find for the first term within
brackets of \eq{done}
\begin{eqnarray}\label{part2}
u^\mu {K^\rho}_\alpha \frac{D u_\rho}{D\tau}\stackrel{\textrm{eff}}{=}\gamma v t^\mu\left[\frac{D u_\alpha}{D\tau}\right]_\perp
.
\end{eqnarray}
Now use \eq{part1} and \eq{part2} in \eq{done}. Shorten the $t^\mu$
components by a $\gamma$ factor (according to the effect of the
inverse), and neglect the $\eta^\mu$-term. We readily find
\begin{eqnarray}\label{kott0}
{P^\mu}_\alpha \frac{D \bar{S}^\alpha}{D\tau}=\frac{\gamma v
}{\gamma+1} \bar{S}^\alpha \left( t^\mu
\left[\frac{D}{D\tau}\left(u_\alpha + \eta_\alpha \right)\right]_\perp 
-t_\alpha \left[ \frac{D}{D\tau}\left(u^\mu + \eta^\mu \right)\right]_\perp  \right)
.
\end{eqnarray}
Now that we have this compact expression we may also find the
$\eta^\mu$ term that we earlier omitted. From orthogonality, $\bar{S}^\alpha
\eta_\alpha=0$, follows that $\frac{D \bar{S}^\alpha}{D \tau}
\eta_\alpha=-\bar{S}^\alpha \frac{D\eta_\alpha}{D\tau}$ which gives
\begin{eqnarray}\label{kott1extra}
\frac{D \bar{S}^\mu}{D\tau}=&&\frac{\gamma v
}{\gamma+1} \bar{S}^\alpha \left( t^\mu
\left[\frac{D}{D\tau}\left(u_\alpha + \eta_\alpha \right)\right]_\perp 
-t_\alpha \left[ \frac{D}{D\tau}\left(u^\mu + \eta^\mu
\right)\right]_\perp  \right) \nonumber\\&&
+\eta^\mu \bar{S}^\alpha \frac{D\eta_\alpha}{D\tau}
.
\end{eqnarray}
So here we have the transport equation for the stopped spin
vector (giving the rotation relative to inertial coordinates).

\section{A note concerning the intrinsic angular momentum}\label{spinapp}
As an idealized scenario we consider a  special relativistic
gyroscope which we model as
an isolated system of point
particles undergoing four-momentum conserving internal
collisions. Following the discussion in \cite{rindler} p. 87-90, 
we define the angular momentum tensor with respect to the
spacetime origin as 
\begin{eqnarray}
L^{\mu \nu}=\sum x^\mu p^\nu - x^\nu  p^\mu. 
\end{eqnarray}
Here the summation runs over 
events $x^\mu$ and four-momenta $p^\mu$ 
for the various particles at
a particular time slice
  $t=const$. The (Pauli-Lubanski) spin vector can be
  written as 
\begin{eqnarray}\label{britt}
S_\mu=\frac{1}{2} \epsilon_{\mu \nu \rho \sigma} L^{\nu \rho} V^\sigma
\end{eqnarray}
Here $V^\mu$ is the four-velocity of the center of
  mass and $\epsilon_{\mu \nu \rho \sigma}$ is the Levi-Civita tensor 
(density) where $\epsilon_{xyz0}=1$. Furthermore
  we introduce an angular momentum four-vector $h^\mu:=(0,{\bf
  h})$, where ${\bf h}$ is the standard (relativistic) angular momentum three-vector, 
  with respect to our reference coordinates. Defining $\eta^\mu$ as a
  purely time directed normalized vector with respect to the reference
  coordinates, we can write
\begin{eqnarray}\label{bratt}
h_\mu=\frac{1}{2} \epsilon_{\mu \nu \rho \sigma} L^{\nu  \rho} \eta^\sigma.
\end{eqnarray}
Letting ${\bf v}$ be the velocity of the center
  of mass, $\gamma$ the corresponding gamma factor and setting 
$(0,{\bf v}):=v t^\mu$ with respect to the reference coordinates, 
we can decompose the four-velocity of the
  center of mass as $V^\mu=\gamma(\eta^\mu+ v t^\mu)$. Using this in
  \eq{britt} together with \eq{bratt}, it follows that 
\begin{eqnarray}
S_\mu=\gamma h_\mu + \gamma v \frac{1}{2}
  \epsilon_{\mu \nu \rho \sigma} L^{\nu  \rho} t^\sigma.
\end{eqnarray}
 It is a short exercise to show that in three-formalism this amounts 
 to 
\begin{eqnarray}
{\bf h}={\bf S}/\gamma + {\bf r}\su{c}\times {\bf p}.
\end{eqnarray}
Here ${\bf h}$ is the net angular momentum of the system of point
particles,  ${\bf r}\su{c}$ is the center of mass (center of energy),
$\gamma$ is the gamma factor for
  the velocity of the center of mass, ${\bf p}$ is the net
  relativistic three-momentum and ${\bf S}$ is the spatial part of the
  spin vector. Note that the intrinsic angular
  momentum is not given by ${\bf S}$ but by ${\bf
  S}/\gamma$. Note incidentally also that there is a difference between the
  center of mass and the 
  {\it proper} center of mass (see \cite{rindler} p. 87-90). As pointed out e.g. in \cite{muller}, the gyroscope
  center of mass does not in general lie on the gyroscope central axis.

  A real gyroscope moving under the influence of forces is
  neither (simply) consisting of point particles nor is it
  isolated. A more detailed analysis would likely assume a general energy
  momentum tensor $T^{\mu \nu}$ and allow for external forces acting on
  the elements of the gyroscope. For the purposes of this article the
  simple derivation outlined above will however suffice.
\\
\\
{\bf References}
\\


\begin{thebibliography}{999}

\bibitem{gravitation}Misner CW, Thorne K S and Wheeler
J A 1973 {\it Gravitation} (New York: Freeman)

\bibitem{intu}Jonsson R 2007 {Gyroscope precession in special and general relativity from basic 
principles}  {\it Am. Journ. Phys.} {\bf 75} 463-471

\bibitem{rickinert}Jonsson R 2006 {Inertial forces and the foundations of optical geometry}
{\it Class. Quantum Grav.} {\bf 23} 1-36

\bibitem{jantzen}Jantzen R T, Carini P and Bini D 1992 {\it
Ann. Phys.} {\bf 215} 1-50 

\bibitem{genopt}Jonsson R and Westman H 2006 {Generalizing optical geometry} {\it Class. Quantum Grav.} 
{\bf 23} 61-76

\bibitem{rindper}Rindler W and Perlick V 1990 {\it Gen. Rel. Grav.}
{\bf 22} 1067-1081

\bibitem{jantzen2}Bini D, Carini P and Jantzen RT 1997
{\it Int. Journ. Mod. Phys. D} {\bf 6} 1-38

\bibitem{jantzen2app}Bini D, Carini P and Jantzen RT 1997
{\it Int. Journ. Mod. Phys. D} {\bf 6} 143-198

\bibitem{rindler}Rindler W (1991) {\it Introduction to special
  relativity} (Oxford: Oxford University Press)

\bibitem{muller}Muller R A 1992 {\it Am. Journ. Phys.} {\bf 60}, 313

\end{thebibliography}
\end{document}